\documentclass[aps,prl,floatfix,showpacs,twocolumn,tightenlines,superscriptaddress]{revtex4}

\usepackage{amsmath}
\usepackage{graphicx}
\usepackage{epsfig}

\begin{document}

\title{A near deterministic linear optical CNOT gate}

\author{Kae Nemoto}\email{nemoto@nii.ac.jp}
\affiliation{National Institute of Informatics, 2-1-2 Hitotsubashi, Chiyoda-ku, Tokyo 101-8430, Japan}

\author{W. J. Munro}\email{bill.munro@hp.com}
\affiliation{National Institute of Informatics, 2-1-2 Hitotsubashi, Chiyoda-ku, Tokyo 101-8430, Japan}
\affiliation{Hewlett-Packard Laboratories, Filton Road, Stoke Gifford,
Bristol BS34 8QZ, United Kingdom}

\date{\today}

\begin{abstract}
We show how to construct a near deterministic CNOT using several single 
photons sources, linear optics, photon number resolving quantum non-demolition 
detectors and feed-forward. This gate does not require the use of massively 
entangled states common to other implementations and is very efficient on 
resources with only one ancilla photon required. The key element of this 
gate are non-demolition detectors that use a weak cross-Kerr nonlinearity 
effect to conditionally generate a phase shift on a coherent probe, if a 
photon is present in the signal mode. These potential phase shifts can 
then be measured using highly efficient homodyne detection. 
\end{abstract}


\maketitle

In the past few  years we have seen the emergence of single photon optics 
with polarisation states as a realistic path for achieving universal 
quantum computation. This started with the pioneering work of Knill, 
Laflamme and Milburn [KLM]\cite{KLM} who showed that with only single 
photon sources and detectors and linear elements such as beam-splitters, 
a near deterministic CNOT gate could be created, through with the use 
of significant but polynomial resources. With this architecture for 
the CNOT gate and trivial single qubit rotations a universal set of 
gates is hence possible and a route forward for creating large devices 
can be seen. Since this original work there has been significant progress 
both theoretically\cite{Pittman01,Knill02,Knill03,Nielsen04,Browne04} and 
experimentally\cite{Pittman03,OBrien03,Gasparoni04}, with a number of 
CNOT gates actually demonstrated.

Much of the theoretical effort has focused on determining more efficient 
ways to perform the controlled logic. The standard model for linear 
logic uses only\cite{KLM}:
\vspace{-2mm}
\begin{itemize}
\item Single photon sources, \vspace{-3mm}
\item Linear optical elements including feed-forward, \vspace{-3mm}
\item Photon number resolving single photon detectors, \vspace{-2mm}
\end{itemize}
and it has been shown by Knill\cite{Knill03} that the maximum probability 
for achieving the CNOT gate is 3/4. While these upper bounds are not 
thought to be tight, with the best success probabilities for the CNOT 
gate being 2/27\cite{Knill01}, it does indicate that near deterministic 
gates are not possible using only the above resources and strategy. These 
gates can be made efficient using the ''standard'' optical teleportation 
tricks which require the use of massively entangled resources. Are there 
other natural ways to increase the efficient of these gate operations? 
Franson {\it et al.}\cite{Pittman01} showed that if you can increase 
your allowed physical resources to include maximally entangled two 
photon states, then the CNOT gate can have its probability of success 
boosted to 1/4, though this is still far below the 3/4 maximum. Alternatively 
it is possible to use single photons for  the cluster state method of one way 
quantum computation\cite{Nielsen04,Browne04}. This can dramatically decrease 
the number of single photons sources required to perform a CNOT gate (from up 
to 10000 for KLM logic to 45 for the cluster approaches). The overhead here 
in single photon sources is large (but polynomial and hence still efficient 
in a sense). Can we however build near deterministic (or deterministic) 
linear optics gates with a low overhead for sources and detectors by 
relaxing the constraints in the standard model? 

There are several options here: we can change the way in which we encode 
our information (from polarisation encoded single photon qubits) or the 
mechanism by which we condition and detect them. There have been schemes 
by Yoran and Reznik\cite{Yoran03} that encode there information in both 
polarisation and which path. This encoding allows a deterministic Bell 
state measurement but the basic gate operations are still relatively 
inefficient. Alternatively one could encode the information in coherent 
states of light as proposed by Ralph et. al\cite{Ralph03}. A key issue 
here becomes the creation and detection of superpositions of coherent 
states. If we want to maintain encoding our information in polarisation 
states of light, what else is possible? The main architecture freedom we 
have left to change are the single photon detectors. We could move to 
nondestructive quantum non-demolition detectors (QND) which would have 
the potential available of be able to condition the evolution of our 
system but without necessarily destroying the single 
photons\cite{Milburn84,Yamamoto85,Grangier98}. They can also resolve one 
photon from a superposition of zero and two. QND devices are generally 
based on cross-Kerr nonlinearities. Historically these reversible 
nonlinearities have been extremely tiny and unsuitable for single 
photon interactions but recently giant Kerr nonlinearities have become 
available with electromagnetically induced transparency (EIT)\cite{Imamoglu96}. 
It is currently not clear whether these nonlinearities are sufficient from the 
natural implementation of single photon-single photon quantum gates, 
however they can be used for QND detection where we require a single 
photon- large coherent beam interaction. Here the nonlinearity strength 
needs to be sufficient only for a small phase shift to be induced onto 
a coherent probe beam (which is distinguishable from the 
original probe)\cite{Munro03}. 

Now that we have decided to use QND detection for linear optical quantum 
computation we need to investigate its effect on the CNOT gates and this 
is the key purpose of this paper. We could investigate each of the known 
gates in turn but we will focus on the FRANSON's 3 photon CNOT 
gate\cite{Pittman03}, the reason being that it requires fewer physical 
detectors to condition the results\footnote{Our results generalise to 
most of the other linear logic cnot gates known. The franson four 
photon gate follows most naturally.}. We will show that a near deterministic 
CNOT gate can be performed with such QND detectors without destroying the 
ancilla photon provided feed-forward is available. More generally we will 
show that for a $n$ qubit circuit, the number of single photon sources 
requires scales as $n+1$. The extra photon is however not destroyed in 
the computation and is left at the end. It is not consumed in the 
computation. This approach can also be applied to achieve cluster 
state computing or computing by measurement alone\cite{Nielsen04,Browne04}. 

Before we begin our detailed discussion, let us first consider the 
photon number QND measurement using a cross-Kerr nonlinearity, which 
has a Hamiltonian of the form $H_{QND}= \hbar \chi a_s^\dagger a_s a_p^\dagger a_p$ 
where the signal (probe) mode has the creation and destruction operators 
given by $a_s^\dagger, a_s$ ($a_p^\dagger, a_p$) respectively and 
$\chi$ is the strength of the nonlinearity. If we consider the signal 
state to have the form $|\psi\rangle= c_0 |0 \rangle_s + c_1 |1 \rangle_s $ 
with the probe beam initially in a coherent state $|\alpha\rangle_p$ then 
the cross-Kerr interaction causes the combined signal/probe system to 
evolve as
\begin{eqnarray}
U_{ck} |\psi\rangle_s |\alpha\rangle_p &=& e^{i H_{QND}t/\hbar} 
\left[c_0 |0 \rangle_s + c_1 |1 \rangle_s\right] |\alpha\rangle_p \nonumber \\
&=& c_0 |0 \rangle_s |\alpha\rangle_p + c_1  |1 \rangle_s |\alpha e^{i \theta }\rangle_p
\end{eqnarray}
where $\theta=\chi t$ with $t$ being the interaction time. We observe immediately 
that the Fock state $|n_a\rangle$ is unaffected by the interaction but the 
coherent state $|\alpha_c\rangle$ picks up a phase shift directly proportional 
to the number of photons $n_a$ in the $|n_a\rangle$ state. For $n_a$ photons 
in the signal mode, the probe beam evolves to $|\alpha e^{i n_a \theta }\rangle_p$. 
Assuming $\alpha \theta \gg 1$ a measurement of the phase of the probe beam (via 
homodyne/heterdyne techniques) projects the signal mode into a definite number 
state or superposition of number states. The requirement $\alpha \theta \gg 1$ 
is interesting as it tells us that a large nonlinearity $\theta$ is not absolutely 
required to distinguish different  $|n_a\rangle$, even for zero, one and two Fock 
states. We could have $\theta$ small but would then require $\alpha$, the amplitude 
of the probe beam large. This is entirely possible and means that we can operate in 
the regime $\theta \ll 1$ which is experimentally more realizable. If this cross-Kerr 
nonlinearity were going to be used directly to implement a CPhase/CNOT gate between 
single photons then we would require $\theta=\pi$. 

In this Fock state detection model we measure the phase of the probe beam 
immediately after it has interacted with the weak cross-Kerr nonlinearity. 
This is the regime where the QND detector functions like the standard single 
photon detector. However, if we want to do a more ''generalised'' type of 
measurement between different signal beams, we could delay the measurement 
of the probe beam instead having the probe beam interact with several 
cross-Kerr nonlinearities where the signal mode is different in each case. The probe 
beam measurement then occurs after all these interactions in a collective way 
which could for instance allow a nondestructive detection that distinguishes 
superpositions and mixtures of the states $|HH\rangle$ and $|VV\rangle$ from 
$|HV\rangle$ and $|VH\rangle$. The key here is that we could have no nett 
phase shifts on the $|HH\rangle$ and $|VV\rangle$ terms while having a 
phase shift on the $|HV\rangle$ and $|VH\rangle$ terms. We will call 
this generalization a {\it two qubit polarisation parity QND detector} 
and it is this type of detector that allows us to circumvent the Knill bounds. 

\begin{figure}[!htb]
\includegraphics[scale=0.37]{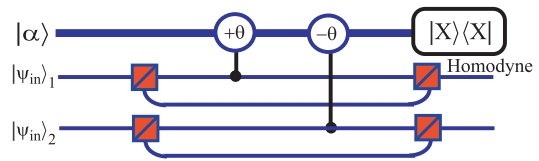} 
\caption{Schematic diagram of a two qubit polarisation QND detector that 
distinguishes superpositions and mixtures of the states $|HH\rangle$ and 
$|VV\rangle$ from $|HV\rangle$ and $|VH\rangle$ using several cross-Kerr 
nonlinearities nonlinearities and a coherent laser probe beam $|\alpha\rangle$. 
The scheme works by first splitting each polarisation qubit into a which path 
qubit on a polarising beamsplitter. The action of the first cross-Kerr nonlinearity 
puts a phase shift $\theta$ on  
the probe beam only if a photon was present in that mode. The second 
cross-Kerr nonlinearity put a phase shift $-\theta$ on the probe beam only if 
a photon was present in that mode. 
After the nonlinear interactions the which 
path qubit are converted back to polarisation encoded qubits. The probe beam 
only picks up a phase shift if the states $|HV\rangle$ and/or $|VH\rangle$ were 
present and hence the appropriate homodyne measurement allows the states 
$|HH\rangle$ and $|VV\rangle$ to be distinguished from $|HV\rangle$ and 
$|VH\rangle$. The two qubit polarisation QND detector thus acts 
like a parity checking device.}
\label{fig-qnd-parity}
\end{figure}

Consider two polarisation qubits initially prepared in the states 
$|\Psi_1\rangle = c_0 |H \rangle_a+ c_1 |V \rangle_a$ and 
$|\Psi_2\rangle = d_0 |H \rangle_b+ d_1 |V \rangle_b$. These qubits 
are split individually on polarizing beam-splitters (PBS) into spatial 
modes which then interact with cross-Kerr nonlinearities as shown in 
Figure (\ref{fig-qnd-parity}).  The action of the PBS's and cross-Kerr 
nonlinearities evolve the combined system 
$|\Psi_1\rangle|\Psi_2\rangle |\alpha\rangle_p $ will evolve to 
$|\psi\rangle_{T}=\left[c_0  d_0 |H H \rangle +c_1  d_1|V V \rangle \right] |\alpha\rangle_p 
+c_0  d_1 |H  V \rangle |\alpha e^{i \theta}\rangle_p+c_1 d_0 |V H \rangle |\alpha e^{-i \theta}\rangle_p$. 
We observe immediately that the $|H H \rangle$ and $|V V \rangle$ pick 
up no phase shift and remain coherent with respect to each other. 
The $|H  V \rangle$ and $|V  H \rangle$ pick up opposite sign phase 
shift $\theta$ which could allow them to be distinguished by a general 
homodyne/heterodyne measurement. However if we choose the local 
oscillator phase $\pi/2$ offset from the probe phase (we will 
call this an X quadrature measurement), then the states 
$|\alpha e^{\pm i \theta}\rangle_p$ can not be distinguished\cite{barrett04}. 
More specifically with $\alpha$ real an $X$ homodyne measurement 
conditions $|\psi\rangle_{T}$ to 
\begin{eqnarray}
&&|\psi_X\rangle_{T}={\it f}(X,\alpha)\left[c_0  d_0 |H H \rangle +c_1  d_1|V V \rangle \right]  \\
&&\;+ {\it f}(X,\alpha cos \theta) \left[ c_0  d_1 e^{i \phi(X)} |H  V \rangle+c_1 d_0 e^{-i \phi(X)}|V  H \rangle \right] \nonumber
\end{eqnarray}
where ${\it f}(x,\beta)=\exp \left[-\frac{1}{4} \left(x-2\beta\right)^2\right]/(2 \pi)^{1/4}$ 
and $\phi(X)= \alpha x \sin \theta -\alpha^2 \sin 2\theta ({\rm Mod} 2 \pi)$. 
We see that ${\it f}(X,\alpha)$ and ${\it f}(X,\alpha cos \theta)$ are 
two Gaussian  curves with the mid point between the peaks located at 
$X_0=\alpha \left[1+\cos \theta \right]$ and the peaks separated by 
a distance $X_d=2 \alpha \left[1-\cos  \theta \right]$. As long as 
this difference is large $\alpha \theta^2 \gg 1$, then there is 
little overlap between these curves. Hence for $X>X_0$ we have
\begin{eqnarray}\label{even-parity}
|\psi_{X>X_0}\rangle_{T}\sim c_0  d_0 |H H \rangle +c_1  d_1|V  V \rangle
\end{eqnarray}
while for $X<X_0$
\begin{eqnarray}\label{odd-parity}
|\psi_{X<X_0}\rangle_{T}\sim c_0  d_1 e^{i \phi(X)} |H  V \rangle+c_1 d_0 e^{-i \phi(X)}|V H \rangle 
\end{eqnarray}
We have used the approximate symbol $\sim$ in these equation as 
there is a small but finite probability that the state 
(\ref{even-parity}) can occur for $X<X_0$. The probability of 
this error occurring is given by  $P_{\rm error}=\frac{1}{2}\left(1-Erf[X_d/{2\sqrt 2}]\right)$ 
which is less than $10^{-5}$ when the distance $X_d \sim \alpha \theta^2 > 9$. 
This shows that it is still possible to operate in the regime of weak 
cross-Kerr nonlinearities, $\theta \ll \pi$. 

The action of this two mode polarisation non-demolition parity detector is 
now very clear; it splits the even parity terms (\ref{even-parity}) nearly 
deterministically from the odd parity cases (\ref{odd-parity}). This is 
really the power enabled by non-demolition measurements and why we can 
engineer strong nonlinear interactions using weak cross-Kerr effects. 
Above we have chosen to call the even parity state \{$|HH\rangle, |VV\rangle$\} 
and the odd parity states \{$|HV\rangle, |VH\rangle$\}, but this is 
an arbitrary choice primarily dependent on the form/type of PBS 
used to convert the polarisation encoded qubits to which path 
encoded qubits. Any other choice is also acceptable and it does 
not have to be symmetric between the two qubits.
 
\begin{figure}[!htb]
\includegraphics[scale=0.29]{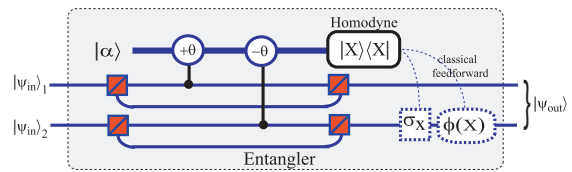}
\caption{Schematic diagram of a two polarisation qubit entangling gate. 
The basis of the scheme uses the QND-based parity detector described in 
Fig (\ref{fig-qnd-parity}). If we consider that the input state of the 
two polarisation qubit is $|HH\rangle+|HV\rangle+|VH\rangle+|VV\rangle$ 
then after the parity gate we have conditioned on an $X$ homodyne 
measurement either the state $|HH\rangle+|VV\rangle$ or 
$e^{i \phi (X)} |HV\rangle+e^{-i \phi (X)}|VH\rangle$ where $\phi(X)$ 
is a phase shift dependent on the result of the homodyne measurement. 
A simple phase shift achieved via classical feed-forward then allows 
this second state to be transformed to the first.}
\label{fig-qnd-entangler}
\end{figure}

It is also interesting to look at the $X<X_0$ solution given 
by (\ref{odd-parity}). We observe immediately that this state is 
dependent on the measured $X$ homodyne value and hence the state 
is conditioned dependent on our measurement result $X$. However 
simple local rotations using phase shifters dependent on the measurement 
result $X$ can be performed via a feed forward process to transform 
this state to $c_0  d_1 |H \rangle_a |V \rangle_b+c_1 d_0 |V \rangle_a |H \rangle_b$ 
which is independent of $X$.  These transformations are very 
interesting as it seems possible with the appropriate choice 
of $c_0, c_1$ and $d_0, d_1$ to create arbitrary entangled 
states near deterministically. For instance if we choose 
$d_0=d_1=1/\sqrt{2}$, then our device outputs either the 
state $c_0 |HH \rangle +c_1|V V \rangle$ or  $c_0|HV \rangle+c_1 |VH \rangle$. 
A simple bit flip on the second polarisation qubit transforms it 
into the first. Thus our two mode parity QND detector can be 
configured to acts as a near deterministic entangler 
(see figure \ref{fig-qnd-entangler}). 
This gate allows us to take two separable polarisation qubits and 
efficiently entangle them (near deterministically). If each of our 
qubits are initially $|H \rangle+ |V \rangle$ then the action of this 
entangling gate is to create the maximally entangled state 
$|HH \rangle +|VV \rangle$. Generally it was thought that 
strong nonlinearities are required to do this near deterministically, 
however our scheme here is using only weak nonlinearities 
$\theta \ll \pi$. This gate is critical and forms the key element 
for our efficient Franson CNOT gate. It can also obviously be used to 
generate maximally entangled state required for several of the 
other CNOT implementations.

\begin{figure}[!htb]
\includegraphics[scale=0.34]{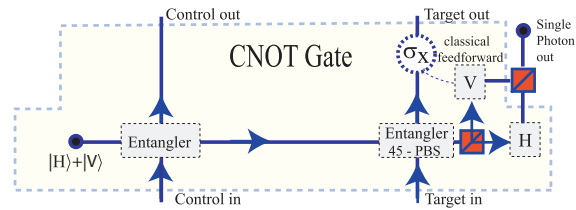}
\caption{Schematic diagram of a near deterministic CNOT composed 
two polarisation qubit entangling gates (one with PBS in 
the \{H,V\} basis and one with PBS in the \{H+V,H-V\} basis), 
one signal photon ancilla prepared as $|H\rangle+|V\rangle$, 
feed-forward elements and four single photon resolving 
QND detectors.}
\label{fig-qnd-franson}
\end{figure}

Now let us move our attention to the construction of the CNOT 
gate (depicted in Fig \ref{fig-qnd-franson}). This is the analogue 
of the Franson CNOT gate from \cite{Pittman03} but with the key 
PBS and 45-PBS replaced with \{H,V\} and $\{D=H+V,\bar D=H-V\}$ 
two polarisation qubit entangling gates. Franson's photon number 
resolving detectors have also been replaced with single photon 
number resolving QND detectors.
Consider an initial state of the form 
$\left[ c_0 |H \rangle_c + c_1 |V \rangle_c \right]\otimes \left[|H\rangle +|V \rangle\right]
\otimes \left[d_0 |H \rangle_t + d_1 |V \rangle_t\right]$. 
The action of the left hand side entangler evolves the system to 
\begin{eqnarray}
\left[c_0 |HH \rangle + c_1 |VV \rangle\right]\otimes \left[d_0 |H \rangle_t + d_1 |V \rangle_t\right]
\end{eqnarray}
Now the action of the 45-entangling gate (where the PBS in the original gate 
have been replaced with 45-PBS's) transforms the state to 
$\left\{c_0 |H \rangle - c_1 |V \rangle\right\} (d_0-d_1)|\bar D,\bar D\rangle+
\left\{c_0 |H \rangle + c_1 |V \rangle\right\} (d_0+d_1)|D,D\rangle$ where for the $X<X_0$ 
measurement we have performed the  usual phase correction, bit flip and an addition 
sign change $|V \rangle\rightarrow -|V \rangle$ on the first qubit). The second 
mode is now split on a normal \{H,V\} PBS and a QND photon number 
measurement performed. A bit flip is performed if a photon is detected 
in the $V$ mode. The final state from these interactions and feed forward 
operations\footnote{There are feedforward operations both in the entangling 
gate and the final measurement step. These can be delayed and performed 
together at the end of the gate} is
\begin{eqnarray}
c_0 d_0 |HH \rangle+c_0 d_1 |HV \rangle+c_1 d_0 |VV \rangle+c_1 d_1 |VH \rangle,
\end{eqnarray} 
which is the same state obtained by performing a CNOT operation on 
the state $\left[ c_0 |H \rangle_c + c_1 |V \rangle_c \right]\otimes 
\left[d_0 |H \rangle_t + d_1 |V \rangle_t\right]$. This shows that 
our QND-based gates has performed a near deterministic CNOT operation. 
The core element of this gate is the {\it two qubit polarisation parity 
QND detector} which engineers a two polarisation qubit interaction 
via a strong probe beam. At the heart of this detector are weak 
cross-Kerr nonlinearities  that make it possible to distinguish 
subspaces of basis states from others which is not possible with 
convenient destructive photon counters. It is this that allows us 
to exceed the Knill bounds presented in \cite{Knill03}. From a 
different perceptive our two mode QND entangling gate is acting 
like a fermonic polarizing beam-splitter, that is it does not 
allow the photon bunching effects. Without these photon bunching 
effects simple feed-forward operations allows our overall CNOT 
gate to be made near deterministic.  This represents a huge 
saving in the physical resources to implement single photon 
quantum logic. For the CNOT operation, only one extra ancilla 
photon is needed beyond the control and target photons to 
perform the gate operation in the near deterministic fashion. 
In fact it is straighforward to observe that if we want to do 
an $n$ qubit computation (with number of one and two qubit gates), 
only $n+1$ single photon sources would be required in principle. 

The resources required to perform this QND based CNOT gate as 
presented here are: three single photon sources, two to encode 
the control and target qubits and one ancilla, six weak 
cross-Kerr nonlinearities, two coherent light laser probe beams 
and homodyne detectors plus basic linear optics elements to 
convert polarisation encoded qubits to spatial coding ones 
and perform the feed-forward. The single photon sncilla is 
not consumed in the gate operation and can be recycled for 
further use. This compares with potentially thousands of 
single photon sources, detectors and linear optical elements 
to implement the original KLM gate. It is possible to construct 
this near deterministic CNOT with fewer cross-Kerr nonlinearities 
(potentially as few as two but recylcing them) but as a cost of 
more feed-forward operations. Finally we should discuss the 
size of the weak cross-Kerr nonlinearity required. Previously 
we have specified a constraint that $\alpha \theta^2 \gg 1$. 
Thus for realistic pumps with mean photon number on the order 
of $10^{12}$ a weak nonlinearity of the order of $\theta=10^{-3}$ 
could be sufficient. While this is still a technological 
challenge it is likely to be achievable in the near future 
and really shows the potential power of weak (but not tiny) 
cross-Kerr nonlinearities. Strong nonlinearities are not a 
prerequisite to be able to perform quantum computation.

{\it To summarize}, We have shown in this letter that weak 
cross-Kerr nonlinearities can be used to construct near 
deterministic CNOT gates with far fewer physical resources 
than other linear optical schemes. This has enormous 
implementations for the development of single photon 
quantum computing and information processing using either 
the convienent models or cluster state techniques. It can 
be immediately be applied to optical cluster state computer 
allowing a significant reduction in the physical resources. 
At the core of the scheme are generalised QND detectors that 
allow us one to distinguish subspaces of the basis states, 
rather than all the basis states which occurs with the 
classic photon counters. The strength of the nonlinearities 
required for our gate are orders of magnitude weaker than 
those required to perform CNOT gates naturally between the 
single photons. Such nonlinearities are potentially available 
today using doped optical fibers, cavity QED and EIT. We hope 
this work motivates the search for weak cross-Kerr nonlinearities 
which now have applications beyond for instance single photon 
number resolving detectors.

\noindent
{\em Acknowledgments}: We will like to thank S. Barrett, 
R. Beausoleil, P. Kok and T. Spiller for valuable discussions. 
This work was supported in part by a JSPS research grant and fellowship, 
an Asahi-Glass research grant and the European Project RAMBOQ.

\end{document}